\documentclass[preprint,aps,amsmath,superscriptaddress,nofootinbib,tightenlines,floatfix]{revtex4}
\usepackage{graphicx}
\usepackage{bm}
\usepackage{epsfig}

\newif\ifpdf
\ifx\pdfoutput\undefined
\pdffalse 
\else
\pdfoutput=1 
\pdftrue
\fi

\def\bsigma{\mbox{\boldmath $\sigma$}}

\def\bnslash{\bar n\!\!\!\slash}

\def\OMIT#1{}

\newcommand{\nn}{\nonumber} 

\newcommand{\bn}{{\bar n}}
\newcommand{\bea}{\begin{eqnarray}}
\newcommand{\eea}{\end{eqnarray}}

\newcommand{\bnP}{\bar {\cal P}}
\newcommand{\ppP}{{\cal P}_\perp}

\newcommand{\cP}{{\cal P}}

\newcommand{\mcdot}{\!\cdot\!}

\newcommand{\mups}{M_\Upsilon}

\newcommand{\vect}[1]{\mathbf{#1}}
\newcommand{\bra}[1]{\langle #1\rvert}
\newcommand{\ket}[1]{\lvert #1\rangle}

\newcommand{\abs}[1]{\left\lvert #1\right\rvert}
\newcommand{\mpipi}{m_{\pi\pi}^2}

\DeclareMathOperator{\Tr}{Tr}

\begin{document}

\ifpdf
\DeclareGraphicsExtensions{.pdf, .jpg}
\else
\DeclareGraphicsExtensions{.eps, .jpg,.ps}
\fi


\preprint{\vbox{ \hbox{CALT-68-2530}\hbox{UCSD/PTH 04-20}  }}

\title{Exclusive Radiative Decays of Upsilon in SCET} 

\author{Sean Fleming}
\affiliation{Department of Physics, University of California at San Diego,
      	La Jolla, CA 92093\footnote{Electronic address: spf@coulomb.ucsd.edu}
	\vspace{0.1cm}}
	
\author{Christopher Lee}
\affiliation{California Institute of Technology, Pasadena, CA 91125\footnote{
	Electronic address: leec@theory.caltech.edu}
	\vspace{0.1cm}}
	
\author{Adam K. Leibovich}
\affiliation{Department of Physics and Astronomy, 	University of Pittsburgh,        Pittsburgh, PA 15260\footnote{Electronic address: akl2@pitt.edu}
	\vspace{0.2cm}}

\date{May 18, 2005\\ \vspace{1cm} }


\begin{abstract}

We study exclusive radiative decays of the $\Upsilon$ using soft-collinear effective theory
and non-relativistic QCD.  In contrast to inclusive radiative decays at the endpoint we 
find that color-octet contributions are power suppressed in exclusive decays, and can safely
be neglected, greatly simplifying the analysis. We determine the complete set of Lorentz 
structures that can appear in the SCET Wilson coefficients and match onto them using results
from a previous calculation. We run these coefficients from the scale $\mups$ to the scale
$\Lambda \sim 1 \, \textrm{GeV}$, thereby summing large logarithms. Finally we use our 
results to predict the ratio of branching fractions $B(\Upsilon \to \gamma f_2)/B(J/\psi \to \gamma f_2)$, $B(J/\psi \to \gamma f_2)/B(\psi' \to \gamma f_2)$, and the partial rate for $\Upsilon \to \gamma \pi \pi$. 

\end{abstract}

\maketitle

\newpage
\section{Introduction}

In a recent series of papers the differential decay rate for the decay $\Upsilon \to \gamma X$
has been studied in the ``endpoint" region where the decay products have a large total 
energy of order the $\Upsilon$ mass ($\mups$), and a small total invariant mass squared of  
order $\Lambda \mups$, where $\Lambda \sim 1 \, \textrm{GeV}$ is the typical hadronic 
scale~\cite{Bauer:2001rh,Fleming:2002rv,Fleming:2002sr,Fleming:2004rk,GarciaiTormo:2004jw}. An important
tool in this analysis is the soft-collinear effective theory (SCET) 
\cite{Bauer:2001ew,Bauer:2001yr,Bauer:2001ct,Bauer:2001yt}, which is a systematic 
treatment of the high energy limit of QCD in the framework of effective field theory. Specifically 
SCET is used to describe the highly energetic decay products in the endpoint region. The heavy 
$b$ and $\bar{b}$  quarks  which form the $\Upsilon$ are described by non-relativistic QCD 
(NRQCD)~\cite{Bodwin:1995jh,Luke:2000kz}.

The soft-collinear effective theory is not limited to applications involving inclusive processes. In 
fact SCET has been extensively applied to exclusive decays of $B$ mesons into light
mesons~\cite{Bauer:2001cu,Bauer:2002aj,Chay:2003zp,Chay:2003ju,Chay:2003kb,Mantry:2003uz,Beneke:2003pa,Lange:2003pk,Leibovich:2003tw,Hill:2004if,Feldmann:2004mg}. Here we use
similar techniques to study exclusive radiative decays of the $\Upsilon$. We make use of some of 
the results derived in the analysis of inclusive radiative decays in the endpoint region, but the analysis of exclusive decays is complicated by the existence of two different collinear scales. This necessitates
a two-step matching procedure~\cite{Bauer:2002aj}. In the first step one matches onto $\textrm{SCET}_{\rm I}$ which describes collinear degrees of freedom with typical offshellness of order 
$\sqrt{\Lambda \mups}$, as is appropriate for inclusive decays in the endpoint region as discussed 
above. In the second step $\textrm{SCET}_{\rm I}$ is matched onto $\textrm{SCET}_{\rm II}$, which is
appropriate for exclusive processes since it describes collinear degrees of freedom with typical offshellness of order $\Lambda$.

The analysis of $\Upsilon$ decay is further complicated by the existence of two types of currents: 
those where the $b\bar{b}$ is in a  color-singlet configuration and those where it is in a 
color-octet configuration. The octet operators are higher-order in the combined NRQCD and SCET
power counting, so one might suppose that they can be dropped. However, the octet currents have 
a Wilson coefficient which is order $\sqrt{\alpha_s(\mups)}$ while the singlet current  has a Wilson coefficient 
of order  $\alpha_s(\mups)$. The additional suppression of  the singlet Wilson coefficient is enough
so that both color-octet and color-singlet operators must be included as  contributions to the inclusive 
radiative decay rate in the endpoint region~\cite{Fleming:2002rv,Fleming:2002sr}. 

In this work we show that in exclusive decays the octet currents are truly suppressed 
relative to the singlet current and can be neglected. We then determine the minimal set of 
color-singlet currents which can arise and fix their matching coefficients in $\textrm{SCET}_{\rm I}$.
We run this current to the intermediate collinear scale $\mu_c \sim \sqrt{\Lambda \mups}$ and
match onto $\textrm{SCET}_{\rm II}$. Our expression for the decay rate agrees to leading order in the twist expansion used in previous work in QCD~\cite{Baier:1,Baier:1985wv,Ma:2001tt}. Finally we use our results to make a prediction for the ratio of branching fractions $B(\Upsilon \to \gamma f_2)/B(J/\psi \to \gamma f_2)$, $B(J/\psi \to \gamma f_2)/B(\psi' \to \gamma f_2)$, and analyze the decay $\Upsilon \to \gamma \pi \pi $ in the kinematic regime where the pions are collinear.

\section{Power Counting}

\subsection{Inclusive Decays}

\begin{figure}[t]
\centerline{\includegraphics[width=6.5in]{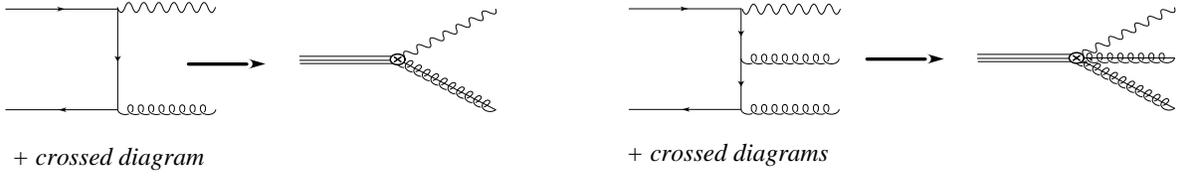}}
\vspace{2em}
\caption{\it Matching onto operators in the effective field theory with one and two gluons in the final state. The currents on the left have a color-octet $b\bar{b}$ in either a ${}^1S_0$ or ${}^3P_J$ configuration. The matching for a color-singlet $b\bar{b}$ pair  in a ${}^3S_1$ configuration is shown on the right.}
\label{matching_fig}
\end{figure}

The first step is to match the QCD amplitude for a $b\bar{b}$ pair in a given color and spin configuration to decay to a photon and light particles onto combined $\textrm{SCET}_{\rm I}$ and NRQCD currents.  The SCET power-counting is in the parameter $\lambda \sim \sqrt{ \Lambda/M}$, where $M = 2m_b$, while the NRQCD power-counting is in $v$, the relative velocity of the $b\bar b$ pair in the $\Upsilon$.  Numerically, $\lambda\sim v \sim1/3$.  The matching is shown graphically in Fig.~\ref{matching_fig}.
The effective theory operators can be classified into those with the $b\bar{b}$ in a color-octet configuration (shown on the left-hand side of Fig.~\ref{matching_fig}) and those with the $b\bar{b}$ in a color-singlet configuration (shown on the right-hand side of Fig.~\ref{matching_fig}). The leading octet operators can be further subdivided into a those where the  $b\bar{b}$ is in a ${}^1S_0$ configuration and those where the $b\bar{b}$ is in a ${}^3P_J$ configuration. The octet ${}^1S_0$ operators are~\cite{Fleming:2002rv,Fleming:2002sr} 
\begin{equation}\label{1s0op2}
J_\mu(8,{}^1S_0) = 
\sum_i C^{8,{}^1S_0}_i(M,\mu) \Gamma^i_{\alpha \mu}\,
 \chi^\dagger_{-{\bf p}} B^\alpha_\perp \psi_{\bf p} \,,
\end{equation}
where
\begin{equation}
B_\perp^\mu = -\frac{i}{g_s}W^\dag[\mathcal{P}_\perp^\mu + g_s(A^\mu_{n,q})_\perp]W.
\end{equation}
The operator $\cP^\mu_\perp$ projects out the label momenta in the perpendicular direction~\cite{Bauer:2001ct}. The sum in Eq.~(\ref{1s0op2}) is over all possible Lorentz structures denoted by $\Gamma^i_{\alpha \mu}$, and
$C^{8,{}^1S_0}_i(M,\mu)$ is the corresponding matching coefficient for each structure. The octet ${}^3P_J$ operators are
\begin{equation} \label{3pjop}
J_\mu(8,{}^3P_J) =
 \sum_i C^{8,{}^3P_J}_i(M,\mu) \Gamma^i_{\alpha \mu \sigma \delta}  
   \chi^\dagger_{-{\bf p}}    B^\alpha_\perp  \Lambda \cdot \widehat{{\bf p}}^\sigma 
   \Lambda \cdot \bsigma^\delta \psi_{\bf p} \,,
\end{equation}
where $\Lambda$ is a Lorentz boost matrix.  Each of these color-octet operators scales as ${\cal O}(\lambda)$ in SCET. The NRQCD power-counting has the ${}^1S_0$ octet operators scaling as ${\cal O}(v^3)$; however, this operator has an overlap with the $\Upsilon$ state beginning at ${\cal O}(v^2)$. Thus the $^1S_0$ operator contributes at order $v^5 \lambda$ to the $\Upsilon$ radiative decay rate. The ${}^3P_J$ octet operator has NRQCD scaling ${\cal O}(v^4)$, but overlaps with the $\Upsilon$ at order $v$. Thus the total power-counting of the $^3P_J$ contribution is ${\cal O}(v^5 \lambda)$, which is the same as the ${}^1S_0$ octet operators. The leading order matching coefficients for both are ${\cal O}(\sqrt{\alpha_s(M)})$. 

The color-singlet operators are 
\begin{equation}\label{3s1op}
J_\mu(1,{}^3S_1) =
\sum_i \Gamma^i_{\alpha \beta \delta \mu}
 \chi^\dagger_{-{\bf p}} \Lambda\cdot\bsigma^\delta \psi_{\bf p}
{\rm Tr} \big\{ B^\alpha_\perp \, 
C^{(1,{}^3S_1)}_i ( M,\bnP_{+} ) \, 
B^\beta_\perp \big\} \,,
\end{equation}
where $\bnP_+ = \bnP^\dagger + \bnP$, with $\bnP \equiv \bn\cdot\cP$. 
These operators scale as ${\cal O}(\lambda^2)$ in SCET and ${\cal O}(v^3)$ in NRQCD. The leading matching coefficients are ${\cal O}(\alpha_s(M))$. Thus the ratio of color-octet to color-singlet contributions in inclusive radiative $\Upsilon$ decay scales as:
\begin{equation}\label{incrat}
\frac{\textrm{octet}}{\textrm{singlet}} \sim 
\frac{v^2}{\lambda \sqrt{\alpha_s(M)}} \sim \frac{v}{\sqrt{\alpha_s(M)}} \,.
\end{equation}

\subsection{Exclusive decays}

The situation changes when one considers exclusive decays. The currents we just discussed are $\textrm{SCET}_{\rm I}$ currents where the typical invariant mass of the collinear degrees of freedom is of order $\mu_c = \sqrt{M \Lambda}$. In order to have overlap with the meson state we must match onto $\textrm{SCET}_{\rm II}$ currents where the typical invariant mass of collinear particles is ${\cal O}(\Lambda)$. Furthermore, the interpolating field which annihilates the meson state in $\textrm{SCET}_{\rm II}$ is defined to consist only of collinear fields in a color-singlet configuration~\cite{Bauer:2002aj}. Given these considerations it is simple to match the color-singlet operator in $\textrm{SCET}_{\rm I}$ to an operator of identical form in $\textrm{SCET}_{\rm II}$.  However, the matching of the octet contributions from $\textrm{SCET}_{\rm I}$ onto $\textrm{SCET}_{\rm II}$ is more involved.

\begin{figure}[t]
\centerline{\includegraphics[width=4.0in]{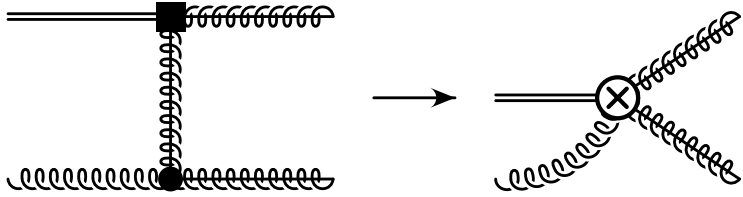}}
\vspace{2em}
\caption{\it An example Feynman diagram of a time-ordered product in $\textrm{SCET}_{\rm I}$ that matches onto an operator in $\textrm{SCET}_{\rm II}$ that has a non-zero overlap with the final state collinear meson.}
\label{octetmatch}
\end{figure}

Before we consider the matching of the octet contributions from $\textrm{SCET}_{\rm I}$ onto $\textrm{SCET}_{\rm II}$ we turn our attention to the scaling of these contributions in $\textrm{SCET}_{\rm I}$. In order to produce a final state consisting only of collinear fields in a color-singlet configuration we need an interaction which changes the ultrasoft (usoft) gluon into a collinear gluon (as shown on the left-hand side of Fig.~\ref{octetmatch}).  This term in the SCET Lagrangian is power suppressed by $\lambda$ so that the time-ordered product of the octet current with the collinear-collinear-usoft vertex scales as ${\cal O}(\lambda^2)$ in the $\textrm{SCET}_{\rm I}$ power counting. In addition, the exchanged gluon introduces an extra factor of the coupling constant at the matching scale: $\alpha_s(\mu_c)$. Including these factors, the time-ordered product of octet currents with the subleading Lagrangian scales as ${\cal O}(\alpha_s(\mu_c) \sqrt{\alpha_s(M)} \lambda^2 v^5)$, and the ratio of time-ordered products to the singlet contribution is
\begin{equation}\label{excrat}
\frac{\textrm{octet}}{\textrm{singlet}} \sim
\frac{v^2 \alpha_s(\mu_c) }{\sqrt{\alpha_s(M)}} \approx 0.05 \,,
\end{equation}
for the bottomonium system. For charmonium the above ratio is about 0.2.
This result is very different from the result for the inclusive decay given in Eq.~(\ref{incrat}). In $\textrm{SCET}_{\rm I}$ the octet contribution to exclusive radiative $\Upsilon$ decay is not only suppressed in the limit $v,\lambda \to 0$, but numerically suppressed by a factor of $\sim 10$ for typical values of the parameters. This is the same order of suppression we expect from higher order SCET and NRQCD corrections; thus, we should be able to safely neglect the color-octet contribution in $\textrm{SCET}_{\rm I}$. However, before we can neglect the octet contribution in our analysis we must show that the suppression of the octet piece holds after matching onto $\textrm{SCET}_{\rm II}$.

We first turn our attention to the simpler calculation: matching the color-singlet operator.  In $\textrm{SCET}_{\rm I}$ we perform the field redefinition \cite{Bauer:2001yt}:
\begin{equation}\label{fieldredef}
A_n \rightarrow Y A^{(0)}_n Y^\dag \,,
\end{equation}
which decouples usoft from collinear degrees of freedom. Under this field redefinition 
\begin{equation}\label{fieldredef2}
B^\alpha_\perp \rightarrow Y B^{(0)\alpha}_\perp Y^\dag\,,
\end{equation}
and the $Y$'s cancel in the trace of the color singlet operator given in Eq.~(\ref{3s1op}). Thus we match the $\textrm{SCET}_{\rm I}$ operator after the field redefinition onto an operator in $\textrm{SCET}_{\rm II}$ of a form identical to that in Eq.~(\ref{3s1op}):
\begin{equation}\label{3s1opII}
J_\mu^{(II)}(1,{}^3S_1) = \sum_i \Gamma^i_{\alpha \beta \delta \mu}
 \chi^\dagger_{-{\bf p}} \Lambda\cdot\bsigma^\delta \psi_{\bf p}
{\rm Tr} \big\{ B^\alpha_{\textrm{II} \perp} \, 
C^{(1,{}^3S_1)}_i ( M,\bnP_{+} ; \mu_c) \, 
B^\beta_{\textrm{II} \perp} \big\} \,.
\end{equation}
where $\mu_c = \sqrt{ M \Lambda}$ is the $\textrm{SCET}_{\rm I}$--$\textrm{SCET}_{\rm II}$ matching scale, and the subscripts indicate $\textrm{SCET}_{\rm II}$ fields. From now on we drop the subscripts.   In $\textrm{SCET}_{\rm II}$ the power-counting parameter is $\eta \sim  \lambda^2$, and the $\textrm{SCET}_{\rm II}$ color-singlet operator in Eq.~(\ref{3s1opII}) is ${\cal O}(v^3 \eta^2)$. The short-distance coefficient is inherited from $\textrm{SCET}_{\rm I}$ and is ${\cal O}(\alpha_s(M))$. 

The matching of the color-octet current is more complicated. In order to match onto an $\textrm{SCET}_{\rm II}$ operator with color-singlet collinear degrees of freedom we must consider time-ordered products where a usoft gluon radiated from the $b\bar{b}$ pair is turned into a final state collinear degree of freedom. An example of such a diagram is given in Fig.~\ref{octetmatch}.
Two collinear gluons are required for the collinear final state to be color-singlet. One of the collinear gluons comes from the octet current, and the other can be produced by pulling a gluon out of the $b\bar b g$ Fock state of the $\Upsilon$, and kicking it with a collinear gluon from the current. This requires a collinear-collinear-ultrasoft coupling which first appears  at order $\lambda$ in the $\textrm{SCET}_{\rm I}$ Lagrangian~\cite{Chay:2002vy,Beneke:2002ph,Pirjol:2002km}:
\begin{equation}
\mathcal{L}_{cg}^{(1)} = \frac{2}{g^2}\Tr\left\{[i\mathcal{D}^\mu, iD_{us}^{\perp\nu}][i\mathcal{D}_\mu, iD_{c\nu}^\perp]\right\},
\end{equation}
where $\mathcal{D}^\mu = D_c^\mu + n\mcdot D_{us}\bar n^\mu/2$. The decay amplitude comes 
from a time-ordered product of the color-octet current and $\mathcal{L}_{cg}^{(1)}$, or a time-ordered product of the color-octet current, $\mathcal{L}_{cg}^{(1)}$, and a leading order gluon interaction. Though our result will hold for either type of time-ordered product we will, for the sake of concreteness, only consider the former:
\begin{equation}\label{top1}
T_8 = \int d^4 x \, T\left\{J(8,\cdot)(0),\mathcal{L}_{cg}^{(1)}(x)\right\},
\end{equation}
where the dot stands for $^1S_0$ or $^3P_J$. In the time-ordered product two gluon fields are contracted to form the internal propagator in Fig.~\ref{octetmatch}, which scales as $1/\lambda^2$. We require two uncontracted $A_{c\nu}^\perp$ fields (in a color-singlet configuration) so that we can match onto an $\textrm{SCET}_{\rm II}$ operator in the form ${\rm Tr}[A_{c\mu}^\perp A_{c\nu}^\perp]$ which annihilates the final state collinear meson. In this example the leading contribution is an $\bar n\cdot A_n$ gluon field in one of the Wilson lines in $J(8,\cdot)$ contracted with an $n\cdot A_n$ field in $\mathcal{L}_{cg}^{(1)}$. After the contraction what remains in $\mathcal{L}_{cg}^{(1)}$ is 
\begin{equation}\label{leftover}
\frac{2}{g^2}\Tr\left\{[gT^A, A_{us}^{\perp\nu}][\bnP, A_{c\nu}^\perp]\right\},
\end{equation}
which scales as $(\lambda^2)(1)(\lambda) = \lambda^3$. Note we now have the correct field content for the operator shown on the right-hand side of Fig.~\ref{octetmatch}: there are two outgoing $A^c_\perp$ fields, one from $\mathcal{L}_{cg}^{(1)}$ and one from $J(8,\cdot )$, in a color-singlet configuration, and an incoming soft gluon field also from $\mathcal{L}_{cg}^{(1)}$.

Next we  decouple collinear and usoft in $\textrm{SCET}_{\rm I}$ through the field redefinition in Eq.~(\ref{fieldredef}). This introduces factors of $Y$ and $Y^\dagger$ into our expressions. When matching onto $\textrm{SCET}_{\rm II}$ these become soft Wilson lines $S$ and $S^\dagger$. Since these Wilson lines do not affect the power counting we ignore them. Now we can match onto a convolution of $\textrm{SCET}_{\rm II}$ operators with $\textrm{SCET}_{\rm I}$--$\textrm{SCET}_{\rm II}$ matching coefficients. Since these arise from integrating out the internal collinear propagators they scale as $\lambda^{-2}$ for each propagator. In our example there is one propagator so the matching coefficient scales as ${\cal O}(\lambda^{-2})$, which is ${\cal O}(\eta^{-1})$ in the $\textrm{SCET}_{\rm II}$ power counting (remember $\eta\sim\lambda^2$). Since the $\textrm{SCET}_{\rm II}$ operator has two $A_{c\nu}^\perp$ fields each scaling as $\eta$, and a soft field scaling as $\eta$, it scales as $\eta^3$. Combining the scaling of the $\textrm{SCET}_{\rm II}$ operator with the scaling of the $\textrm{SCET}_{\rm I}$--$\textrm{SCET}_{\rm II}$ matching coefficient gives an ${\cal O}(\alpha_s(\mu_c) \eta^2)$ contribution. If we include the order $v^5$ NRQCD scaling from the heavy sector, and the ${\cal O}(\sqrt{\alpha_s(M)})$ contribution from the QCD--$\textrm{SCET}_{\rm I}$ matching coefficient the color-octet contribution to exclusive decays scales as ${\cal O}(v^5 \eta^2 \sqrt{\alpha_s(M)} \alpha_s(\mu_c))$ in $\textrm{SCET}_{\rm II}$. Taking the ratio of the color-octet to the color-singlet contribution to exclusive $\Upsilon$ decay in  $\textrm{SCET}_{\rm II}$ we find: 
\begin{equation}
\frac{\textrm{octet}}{\textrm{singlet}} \sim \frac{v^2 \alpha_s(\mu_c)}{\sqrt{\alpha_s(M)}} \,,
\end{equation}
which is the same scaling we found in $\textrm{SCET}_{\rm I}$. Thus we can safely neglect the color-octet contributions at this order.

\section{Complete Basis of Color-Singlet Matching Coefficients}

Now that the color-octet contribution has been eliminated we determine a complete basis of Lorentz structures $\Gamma^i_{\alpha\beta\mu\nu}$  that can appear in the color-singlet matching coefficient in Eq.~(\ref{3s1op}). At leading order in $\alpha_s(M)$ only one Lorentz structure was found to be non-zero~\cite{Fleming:2002rv}:
\begin{equation}\label{matchcoeff1}
C_1^{(1,{}^3S_1)}(M,\omega) \Gamma^1_{\alpha\beta\delta\mu} = 
\frac{4 g^2_s e e_b}{3 M} g^\perp_{\alpha\beta} g_{\mu\delta} \,.
\end{equation}
However, at higher order other Lorentz structures may appear. These coefficients can be constructed from the set:
\begin{equation}\label{set}
\{ g_{\mu\nu}, n_\mu, \bar n_\mu, v_\mu \} \,,
\end{equation}
where $v$ is the four-velocity of the $\Upsilon$, under the restriction that $\Gamma^i_{\alpha\beta\delta\mu}$ satisfies the appropriate symmetries. For example, the full theory amplitude is parity even, as is the effective theory operator, meaning that the matching coefficient must also be parity even. As a result the epsilon tensor is not included in Eq.~(\ref{set}). \

We treat $v$ as an object independent of $n,\bar n$~\cite{Pirjol:2002km}, and use $n^2 = \bar n^2 = 0$, $n\cdot\bar n = 2$, and $v^2 = 1$. Before we write down all the possible operators which can appear we note some simple properties that will make our task more manageable. First we note that $\Gamma_{\alpha\beta\delta\mu}$ must be symmetric in $\alpha$ and $\beta$. To see this consider the object:
\begin{equation}\label{blah}
\sum_\omega C_i^{(1,{}^3S_1)}(M,\omega) \Gamma^i_{\alpha\beta\delta\mu} 
\Tr(B_\perp^\alpha \delta_{\omega, \bnP_{+}}  B_\perp^\beta) \,.
\end{equation}
which is the collinear part of the color-singlet operator where a sum over $\omega$ has been introduced. First interchange $\alpha$ and $\beta$, and then use the cyclic nature of the trace to switch the two $B_\perp$ fields. Note, however, there is a projection on these fields from the operator in the Kronecker delta involving $\bnP_{+}$. Since $\bnP_{+} = \bnP^\dagger + \bnP$ this operator projects out minus the label on $B_\perp^\alpha$ and projects out the label on $B_\perp^\beta$~\cite{Bauer:2001ct}. To preserve this relationship when the order of the fields is switched we must let $\delta_{\omega, \bnP_{+}} \to  \delta_{\omega, -\bnP_{+}}$. By letting $\omega \to -\omega$ we  have $\delta_{-\omega, -\bnP_{+}} =  \delta_{\omega, \bnP_{+}}$, and the operator goes into itself. However, the Wilson coefficient is now $C_i^{(1,{}^3S_1)}(M,-\omega)$. To demonstrate that Eq.~(\ref{blah}) is symmetric under $\alpha \leftrightarrow \beta$ we must show that $C_i^{(1,{}^3S_1)}(M,\omega)$ is even in $\omega$. We use charge conjugation for this. The heavy quark sector of the operator has charge conjugation $C=-1$ as does the photon. As noted in Ref.~\cite{Novikov:1977dq} two gluons in a color-singlet configuration must have $C$ even. Since QCD is charge conjugation conserving the product of operator and coefficient in Eq.~(\ref{blah}) must also be $C$ even. This is the case if the matching coefficient $ C_i^{(1,{}^3S_1)}(M,\omega) \Gamma^i_{\alpha\beta\delta\mu}$ is $C$ even. Following Ref.~\cite{Bauer:2002nz}, under charge conjugation the above product of operator and coefficient goes to itself with $\omega \to -\omega$ in the coefficient function. Thus charge conjugation implies $C_i^{(1,{}^3S_1)}(M,-\omega) = C_i^{(1,{}^3S_1)}(M,\omega)$, and as a result Eq.~(\ref{blah}) is symmetric in $\alpha$ and $\beta$.  
Second, any $v_\delta$ appearing in $\Gamma_{\alpha\beta\delta\mu}$ gives zero contribution to the operator, since $v\cdot\Lambda = 0$. Third, $n_\alpha,\bar n_\alpha$ (and by symmetry $n_\beta, \bar n_\beta$) appearing in the operator also gives a zero contribution since these indices contract with indices on the $B_\perp$ field. Finally, we use reparameterization invariance (RPI) of SCET~\cite{Chay:2002vy,Manohar:2002fd}. The terms satisfying these requirements are
\begin{align}
\label{coeff1}
\sum_i C_i^{(1,{}^3S_1)}(M,\omega) \Gamma^i_{\alpha\beta\delta\mu}  & =
c_1 g_{\alpha\beta}g_{\delta\mu} + c_2 g_{\alpha\beta}\frac{n_\delta n_\mu}{(n\cdot v)^2} + c_3 g_{\alpha\beta}\frac{n_\delta v_\mu}{n\cdot v} \\
& \hspace{-15ex}
+c_4\biggl[\left(g_{\alpha\mu}-\frac{v_\alpha n_\mu}{n\cdot v}\right)\left(g_{\beta\delta}-\frac{v_\beta n_\delta}{n\cdot v}\right) + \left(g_{\alpha\delta}-\frac{v_\alpha n_\delta}{n\cdot v}\right)\left(g_{\beta\mu}-\frac{v_\beta n_\mu}{n\cdot v}\right)\biggr] \,.
\nonumber
\end{align}

So far we have allowed $v$ to be an arbitrary vector. Now we restrict ourselves to a frame where $v_\perp^\mu$ = 0.  Furthermore we are interested in the case where the photon is real, so we can restrict the photon to have transverse polarizations. This leaves only two linearly independent terms:
\begin{equation}
\label{coefftrans}
\sum_i C_i^{(1,{}^3S_1)}(M,\omega) \Gamma^i_{\alpha\beta\delta\mu} = a^g_1 g^\perp_{\alpha\beta}g^\perp_{\delta\mu} + a^g_2\left( g^\perp_{\alpha\delta}g^\perp_{\beta\mu} + g^\perp_{\alpha\mu}g^\perp_{\beta\delta} - g^\perp_{\alpha\beta}g^\perp_{\delta\mu}\right)\,,
\end{equation}
where $g^{\mu\nu}_\perp = g^{\mu\nu}-(n^\mu \bn^\nu + \bn^\mu n^\nu)/2$. The first term projects out the trace part of the $\Tr(B_\perp^\alpha \delta_{\omega, \bnP_{+}}  B_\perp^\beta)$ operator, while the second term projects out the symmetric traceless component. Since the Lorentz symmetry in the perpendicular components is not broken in SCET these two terms do not mix under renormalization. The leading-order matching fixes the coefficients $a_1$ and $a_2$ to order $\alpha_s$:
\begin{equation}\label{matchcoeff}
a^g_1(\bnP_{+} ; \mu = M)  = \frac{4g^2_s e e_b}{3M}, \hspace{10ex}
a^g_2(\bnP_{+} ; \mu = M)  = 0 \,.
\end{equation}
Since there is no mixing, $a_2=0 + {\cal O}(\alpha_s^2(M))$ at all scales. Note this matching assumes the $\Upsilon$ states are non-relativistically normalized: $\langle \Upsilon (P') | \Upsilon (P) \rangle = \delta^3(P-P')$. 

In addition to gluon operators we must consider the basis of all possible quark operators which can appear in radiative $\Upsilon$ decays
\begin{equation}\label{quarkcur}
J_\mu^q (1,{}^3S_1) = \sum_{i} \chi^\dagger_{-{\bf p}} \Lambda\cdot\bsigma^\delta \psi_{\bf p}
\bar{\xi}_{n,p_1} W \Gamma^i_{\mu \delta} (M,\bnP_+) W^\dagger \xi_{n,p_2} \,.
\end{equation}
The basis of Dirac structures, $\{ \bnslash , \bnslash \gamma_5 , \bnslash \gamma^\mu_\perp \}$, was given in Ref.~\cite{Bauer:2002nz}, and the most general basis of quark operators can then be constructed out of these Dirac structures and the set $\{ \epsilon_{\alpha\beta\mu\nu}, g_{\mu\nu}, n_\mu, \bar n_\mu, v_\mu \}$. Using the symmetries of SCET and RPI we find
\begin{equation}
\sum_{i} \Gamma^i_{\mu \delta}  = 
a^q_1 \frac{\bnslash}{2} g^\perp_{\mu\delta} + 
a^q_2  \frac{\bnslash}{2} \gamma_5 \epsilon^\perp_{\mu\delta} + 
a^q_3  \frac{\bnslash}{2} \gamma^\perp_{\mu} n_{\delta} \,.
\end{equation}
The first term transforms as a scalar, the second term transforms as a pseudoscalar, and the third as a vector. 
The matching coefficients at the scale $\mu = M$ for the quark operators in radiative $\Upsilon$ decay are all zero at leading order in perturbation theory, but the scalar quark operator mixes with the scalar gluon operator through renormalization group running and can be generated in this manner. The pseudoscalar and vector term do not mix with the scalar gluon operator due to Lorentz symmetry and will not be generated at this order in the perturbative matching.

\section{Decay Rates \& Phenomenology}

We now consider the phenomenological implications of our analysis for exclusive radiative decays of quarkonium into either a single meson or a pair of mesons which are collinear. The $(n+1)$-body decay rate is given by:
\begin{eqnarray}
\Gamma (\Upsilon \to \gamma F_n ) &=& \frac{1}{2\mups} \int \frac{d^3\vect{q}}{2 E_{\gamma} (2\pi)^3}
\prod^n_i \frac{d^3\vect{p}_i}{2 E_i (2\pi)^3} (2 \pi )^4 \delta^4(P - q - \sum^n_i p_i) 
\nn \\
& & \hspace{10ex} \times
| \langle \gamma(q) p_1 ... p_n | J^\mu\mathcal{A}_\mu | \Upsilon(P) \rangle |^2
\end{eqnarray}
where $J$ is the QCD current, $\mathcal{A}$ is the photon field, and $F_n$ denotes an exclusive final state consisting of $n$ collinear particles. We consider only decay rates where the final state momenta $p_i$ are all collinear with combined invariant mass $m^2_n = (\sum^n_i p_i)^2 \sim \Lambda$. The  effective theory decay rate is obtained by matching the current $J$ onto the $\textrm{SCET}_{\rm II}$ current given in Eq.~(\ref{3s1opII}) plus a quark operator:
\begin{eqnarray}\label{currmatch}
& &  \langle \gamma(q) p_1 ... p_n | J^\mu\mathcal{A}_\mu | \Upsilon(P) \rangle
\nn \\
& \to &
\sum_i \Gamma^i_{\alpha \beta \delta \mu} \langle  \gamma(q)  p_1 ... p_n  | 
 \chi^\dagger_{-{\bf p}} \Lambda\mcdot\bsigma^\delta \psi_{\bf p}
{\rm Tr} \big\{ B^\alpha_{\perp} \, 
C^{(1,{}^3S_1)}_i ( \bnP_{+} ; \mu) \, 
B^\beta_{ \perp} \big\} \mathcal{A}^\mu |  \Upsilon(P) \rangle
\nn \\
& & + \sum_i  \langle  \gamma(q)  p_1 ... p_n  | 
 \chi^\dagger_{-{\bf p}} \Lambda\mcdot\bsigma^\delta \psi_{\bf p}
\bar{\xi}_{n,p_1} W \Gamma^i_{\mu \delta} (M,\bnP_+) W^\dagger \xi_{n,p_2} 
\mathcal{A}^\mu|  \Upsilon(P) \rangle
\nn \\
& = & 
\bra{0}\chi_{-\vect{p}}^\dag\Lambda\mcdot\bsigma^\delta\psi_{\vect{p}}\ket{\Upsilon(P)}
\bra{\gamma(q)}\mathcal{A}^\mu\ket{0}g^\perp_{\delta\mu} \nn \\
& & \times\bigg[ \langle p_1 ... p_n  |  {\rm Tr} \big\{ B^\alpha_{\perp} \, 
a^g_1( \bnP_{+} ; \mu) \, 
B_\alpha^{ \perp} \big\}  | 0 \rangle  + \langle p_1 ... p_n  |  \bar{\xi}_{n,p_1} W \frac{\bnslash}{2} \, \frac{a^q_1( \bnP_{+} ; \mu)}M \,
W^\dagger \xi_{n,p_2} | 0 \rangle 
\bigg] \,.
\end{eqnarray}
In obtaining the second line we make use of the results in Eqs.~(\ref{coefftrans}) and~(\ref{matchcoeff}), and use the properties of $\textrm{SCET}_{\rm II}$ to factor soft and collinear degrees of freedom.   In the last line we  changed to a nonrelativistic normalization for the $\Upsilon$ state.

Next we define the light-cone wave functions
\begin{equation}
\label{wfdefs}
\begin{split}
\bra{p_1\dots p_n} \mathcal{\bar P}\Tr[B_\perp^\alpha \delta_{\omega,\mathcal{P}_+} B^\perp_\alpha]\ket{0} &= M^{3-n} \phi_g^{F_n}(x),  \\
\bra{p_1\dots p_n}\bar\xi_{n,\omega_1} W \frac{\bnslash}{2}\delta_{\omega,\bnP_+} W^\dag\xi_{n,\omega_2} \ket{0} &= M^{3-n}\phi_q^{F_n}(x) \,,
\end{split}
\end{equation}
where states are relativistically normalized, and the discrete label $\omega$ is converted to a continuous one, $x = \omega/\bar{n}\mcdot p$, as explained in Ref.~\cite{Fleming:2004rk}. The wave functions $\phi_{q,g}^{F_n}$ are dimensionless. See Appendix~\ref{app:lc} for the relation of these SCET light-cone wave functions to those conventionally defined in QCD.  Then the collinear matrix elements in brackets in Eq.~(\ref{currmatch}) can be written as the convolution:
\begin{eqnarray}
& & 
\langle p_1 ... p_n  |  {\rm Tr} \big\{ B^\alpha_{\perp} \, 
a^g_1( \bnP_{+} ; \mu) \, 
B_\alpha^{ \perp} \big\}  | 0 \rangle 
 + \langle p_1 ... p_n  |  \bar{\xi}_{n,p_1} W \frac{\bnslash}{2} \, a^q_1( \bnP_{+} ; \mu) \,
W^\dagger \xi_{n,p_2} | 0 \rangle 
\nn \\
& & =  
M^{2-n}\int^1_{-1} dx \, \big( a_1^g(x;\mu)  \phi^{F_n}_g(x;\mu) + a_1^q(x;\mu)  \phi^{F_n}_q(x;\mu) \big)\,.
\end{eqnarray}
The dependence on the scale $\mu$ cancels between the long-distance matching coefficients and the wave function. We will elaborate on this point in a moment. First we expand both $a^{g/q}_1(x;\mu)$ and $\phi^{F_n}_{g/q}(x;\mu)$ in Gegenbauer polynomials:
\begin{eqnarray}
a_1^q(x;\mu) &=& \sum_{n\textrm{ odd}}a_q^{(n)}(\mu)C_n^{3/2}(x) \,,
\nn \\
a_1^g(x;\mu) &=& \sum_{n\textrm{ odd}}a_g^{(n)}(\mu)(1-x^2)C_{n-1}^{5/2}(x) \,,
\nn \\
\phi_q^{F_n}(x;\mu) &=& \sum_{n\textrm{ odd}}b_q^{(n)}(\mu)(1-x^2)C_n^{3/2}(x) \,, 
\nn \\
\phi_g^{F_n}(x;\mu) &=& \sum_{n\textrm{ odd}}b_g^{(n)}(\mu)(1-x^2)C_{n-1}^{5/2}(x) \,.
\end{eqnarray}
Then the convolution becomes an infinite sum of products of Gegenbauer coefficients:
\begin{eqnarray}\label{gegenbauer}
&&  \int^1_{-1} dx \big( a_1^g(x;\mu)  \phi^{F_n}_g(x;\mu) + a_1^q(x;\mu)  \phi^{F_n}_q(x;\mu) \big) 
\nn \\
&& = \sum_{n\textrm{ odd}} \big( f^{(n)}_{5/2} a_g^{(n)}(\mu) b_g^{(n)}(\mu) + 
f^{(n)}_{3/2} a_q^{(n)}(\mu) b_q^{(n)}(\mu)\big) \,,
\end{eqnarray}
where 
\begin{equation}
f^{(n)}_{5/2} = \frac{n(n+1)(n+2)(n+3)}{9(n+3/2)} \,, \hspace{5ex} 
f^{(n)}_{3/2} = \frac{(n+1)(n+2)}{n+3/2} \,.
\end{equation}

We now return to the question of the scale. Here we pick $\mu \sim \Lambda$ which minimizes logarithms in the wave function; however, large logarithms of $M/\Lambda$ then appear in the Wilson coefficients. These large logarithms are summed using the renormalization group equations in SCET. This calculation was carried out in Ref.~\cite{Fleming:2004rk}, and we only quote the results here. We find:
\begin{eqnarray}\label{resum}
&&  \int^1_{-1} dx \big( a_1^g(x;\mu)  \phi^{F_n}_g(x;\mu) + a_1^q(x;\mu)  \phi^{F_n}_q(x;\mu) \big) 
\nn \\
&& = \frac{4}{3} a_1^g(M) \sum_{n\textrm{ odd}} \bigg\{ \bigg[ 
\gamma^{(n)}_+ \bigg( \frac{\alpha_s(\mu)}{\alpha_s(M)} \bigg)^{2 \lambda^{(n)}_+/\beta_0} -
\gamma^{(n)}_- \bigg( \frac{\alpha_s(\mu)}{\alpha_s(M)} \bigg)^{2 \lambda^{(n)}_-/\beta_0} \bigg]
b_g^{(n)}(\mu)
\nn \\
&& \hspace{10ex} + 
\frac{f^{(n)}_{3/2}}{f^{(n)}_{5/2}} \frac{\gamma^{(n)}_{gq}}{\Delta^{(n)}} \bigg[ 
 \bigg( \frac{\alpha_s(\mu)}{\alpha_s(M)} \bigg)^{2 \lambda^{(n)}_+/\beta_0} -
 \bigg( \frac{\alpha_s(\mu)}{\alpha_s(M)} \bigg)^{2 \lambda^{(n)}_-/\beta_0} \bigg]
b_q^{(n)}(\mu) \bigg\} \,, 
\end{eqnarray}
where 
\begin{eqnarray}
\beta_0 &=& 11 - \frac{2n_f}{3} \,, \nn \\
\gamma_\pm^{(n)} &=& \frac{\gamma_{gg}^{(n)} - \lambda_\mp^{(n)}}{\Delta^{(n)}} \,,
\nn \\
\lambda_\pm^{(n)} &=& \frac{1}{2}\left[\gamma_{gg}^{(n)} + \gamma_{q\bar q}^{(n)} \pm \Delta^{(n)}\right] \,,
\nn \\
\Delta^{(n)} &=& \sqrt{\left(\gamma_{gg}^{(n)} - \gamma_{q\bar q}^{(n)}\right)^2 + 4\gamma_{gq}^{(n)}\gamma_{qg}^{(n)}} \,,
\nn \\
\gamma_{q\bar q}^{(n)} &=& C_F\left[\frac{1}{(n+1)(n+2)} - \frac{1}{2} - 2\sum_{i=2}^{n+1}\frac{1}{i}\right]  \,,
\nn \\
\gamma_{gq}^{(n)} &=& \frac{C_F}{3}\frac{n^2+3n+4}{(n+1)(n+2)}  \,,
\nn \\
\gamma_{qg}^{(n)} &=& 3n_f\frac{n^2+3n+4}{n(n+1)(n+2)(n+3)}  \,,
\nn \\
\gamma_{gg}^{(n)} &=& C_A\left[\frac{2}{n(n+1)} + \frac{2}{(n+2)(n+3)} - \frac{1}{6} - 2\sum_{i=2}^{n+1}\frac{1}{i}\right] - \frac{n_f}{3}.
\end{eqnarray}
The quantities $\lambda^{(n)}_\pm$ which appear in the exponents in  Eq.~(\ref{resum}) are negative for any $n >1$. Furthermore $\lambda^{(1)}_- < 0$, while $\lambda^{(1)}_+ = 0$. This property allows us to consider the asymptotic limit $M \gg\Lambda$, where $\alpha_s(M) \to 0$. Then
\begin{equation}
\begin{split}
\lim_{M \to \infty}  \int_{-1}^1 dx
&\left[a_1^g(x;\mu)\phi_g^{F_n}(x;\mu) + a_1^q(x;\mu)\phi_q^{F_n}(x;\mu)\right] \\
&\longrightarrow   
\frac{16}{3} \frac{C_F}{4C_F + n_f} 
a_1^g(M)\left[b_g^{(1)}(\Lambda) + \frac{3}{4}b_q^{(1)}(\Lambda)\right] \equiv B^{F_{n}} a_1^g(M), 
\end{split} 
\end{equation}
which defines a nonperturbative parameter $B^{F_n}$. However, for values of $M$ around the $\Upsilon$ mass this is not a very good approximation, and for values around the $J/\psi$ mass a much better approximation is to assume no running at all.

\subsection{Two body decay: $\Upsilon\rightarrow\gamma f_2$}

Having taken care of the technical details we can now use the above results to study the two body radiative decay $\Upsilon \to \gamma F_1$. The decay rate is
\begin{eqnarray}
\label{2body}
\Gamma (\Upsilon \to \gamma F_1 )_{\textrm{SCET}_{\rm II}} & = & \frac{1}{16\pi} 
\bra{\Upsilon } \psi_{{\bf p}'}^\dagger \sigma_\perp^i \chi_{-{\bf p}'}  
\chi^\dagger_{-{\bf p}} \sigma_\perp^i \psi_{\bf p} \ket{ \Upsilon} 
\nn \\ 
& &  \times  
\bigg[ \int^1_{-1} dx \big( a_1^g(x;\mu)  \phi^{F_1}_g(x;\mu) + a_1^q(x;\mu)  \phi^{F_1}_q(x;\mu) \big) \bigg]^2,
\end{eqnarray}
where the full expression for the term in brackets is given in Eq.~(\ref{resum}). After factoring, the soft matrix element involving the heavy quark fields was further simplified using the vacuum insertion approximation for the quarkonium sector, which holds up to corrections of order $v^4$~\cite{Bodwin:1995jh}. Note that the operator above overlaps only with the $\lambda=\pm 1$ helicities of the $\Upsilon$. Then using the rotation symmetries of NRQCD~\cite{Braaten:1996jt} we can relate the non-relativistic matrix element above to those conventionally used:
\begin{equation}
\bra{\Upsilon } \psi_{{\bf p}'}^\dagger \sigma_\perp^i \chi_{-{\bf p}'}  
\chi^\dagger_{-{\bf p}} \sigma_\perp^i \psi_{\bf p} \ket{ \Upsilon} =
\frac{2}{3} \bra{\Upsilon } {\cal O}(1,{}^3S_1)  \ket{ \Upsilon} \,.
\end{equation}
For the final state meson $F_1$ to have nonzero overlap with the operators in Eq.~(\ref{wfdefs}) it must be flavor singlet, parity even and charge conjugation even. One candidate with the correct quantum numbers is the $f_2(1270)$. Furthermore this decay has been measured both in $\Upsilon$ and $J/\psi$ radiative decay, which is why we consider it. An interesting point is that only the helicity $\lambda =0$ component of the $f_2$ contributes at the order to which we are working. To see this begin
by considering the decomposition of the following gluon matrix element into all possible light-cone form-factors:
\begin{equation}
\label{f2formfact}
\bra{f_2}\Tr[B_\perp^\alpha B_\perp^\beta]\ket{0} = 
A(e(\lambda))g_\perp^{\alpha\beta} + B_\lambda e_\perp^{\alpha\beta}(\lambda),
\end{equation}
where $e^{\alpha\beta}$ is the symmetric-traceless polarization tensor of the $f_2$. We give the explicit form in Appendix~\ref{sec:pol}.  There are only two form factors above since the matrix element must be decomposed into tensors that have non-zero perpendicular components. The only structures available are $g_\perp^{\alpha\beta}$ and $e_\perp^{\alpha\beta}$. For $\lambda=\pm 1$, $e_\perp^{\mu\nu}(\lambda=\pm 1) = 0$, so this helicity component does not appear at this order. The coefficient $A(e(\lambda))$ is a scalar function which can be constructed from $\Tr(e_\perp)$ and $\bn_\alpha n_\beta e^{\alpha\beta}$. Because the helicity-zero polarization tensor has the property that $e_\perp^{\mu\nu}(\lambda = 0)\propto g_\perp^{\mu\nu}$, and the helicity-two polarization tensor has $\Tr(e_\perp) = 0$ and $\bn_\alpha n_\beta e^{\alpha\beta}= 0$, we can fix the normalization of  the coefficient $A(e(\lambda))$ so  that the first  term on the right-hand side of Eq.~(\ref{f2formfact}) parameterizes the $\lambda = 0$ contribution while the second term parameterizes the $\lambda = \pm 2$ contributions. The helicity-zero piece is picked out by the $a_1 g_\perp^{\alpha\beta}$ term in Eq.~(\ref{coefftrans}), while the helicity-two piece is picked out by the $a_2$ term. Thus at leading order in perturbation theory, the dominant decay should be to the helicity-zero component of the $f_2$.

The NRQCD matrix element in Eq.~(\ref{2body}) can be expressed in terms of the leptonic decay width of the $\Upsilon$. At leading order,
\begin{equation}
\Gamma(\Upsilon\rightarrow e^+ e^-) = \frac{8\pi\alpha^2 e_b^2}{3 M^2}\bra{\Upsilon} {\cal O}(1,{}^3S_1)   \ket{ \Upsilon} \,,
\end{equation}
and the decay rate for $\Upsilon\rightarrow\gamma f_2$ can be expressed as:
\begin{equation}
\Gamma(\Upsilon\rightarrow\gamma f_2) = \frac{16\pi\alpha_s(M)^2}{9\alpha}(B^{f_2})^2\Gamma(\Upsilon\rightarrow e^+ e^-)\,.
\end{equation}
We can repeat the same analysis for the decay rate $\Gamma(J/\psi\rightarrow\gamma f_2)$ and form a ratio of branching fractions, which in the asymptotic limit is:
\begin{equation}
\frac{B(\Upsilon\rightarrow\gamma f_2)}{B(J/\psi\rightarrow\gamma f_2)} = \left[\frac{\alpha_s(M_{b\bar{b}})}{\alpha_s(M_{c\bar{c}})}\right]^2\left(\frac{4 C_F +  3}{4C_F+4}\right)^2\frac{B(\Upsilon\rightarrow e^+ e^-)}{B(J/\psi\rightarrow e^+ e^-)}\,,
\end{equation}
where $M_{Q\bar{Q} } = 2 m_Q$.  Using $m_b = 4.1-4.4$ GeV, $m_c = 1.15-1.35$ GeV, $B(\Upsilon\rightarrow e^+e^-) = (2.38\pm 0.11)\times 10^{-2}$, and $B(J/\psi\rightarrow e^+e^-) = (5.93\pm 0.10)\times 10^{-2}$ \cite{Eidelman:2004wy}, we predict the ratio of branching fractions to be in the range:
\begin{equation}
\label{ratioasymptotic}
\left[\frac{B(\Upsilon\rightarrow\gamma f_2)}{B(J/\psi\rightarrow\gamma f_2)}\right]_{M\rightarrow\infty} = 0.14-0.19\,.
\end{equation}
As was mentioned earlier the asymptotic limit is not particularly good for the $\Upsilon$, and quite bad for the $J/\psi$. As a consequence we consider the scenario where the resummation of logarithms is neglected. In this case the ratio of branching fractions lies in the range
\begin{equation}
\label{nologs}
\frac{B(\Upsilon\rightarrow\gamma f_2)}{B(J/\psi\rightarrow\gamma f_2)} = 0.18 - 0.23\,.
\end{equation}
We can improve this approximation by keeping more terms in the resummed formula in Eq.~(\ref{resum}). The dominant term is the part of the $n=1$ term proportional to $b_g^{(1)}(\mu)$, and in this approximation:
\begin{eqnarray}
\label{ratioimproved}
\frac{B(\Upsilon\rightarrow\gamma f_2)}{B(J/\psi\rightarrow\gamma f_2)} &=& \left[\frac{\alpha_s(M_{b\bar{b}})}{\alpha_s(M_{c\bar{c}})}\right]^2
\left[\frac{ \gamma_+^{(1)}  - 
                 \gamma_-^{(1)} \big( \alpha_s(\mu ) / \alpha_s(M_{b\bar{b}}) \big)^{2 \lambda_-^{(1)}/\beta_0^{n_f=4}} }
               { \gamma_+^{(1)}  - 
                 \gamma_-^{(1)} \big( \alpha_s(\mu ) / \alpha_s(M_{c\bar{c}}) \big)^{2 \lambda_-^{(1)}/\beta_0^{n_f=3}} }
\right]^2
\frac{B(\Upsilon\rightarrow e^+ e^-)}{B(J/\psi\rightarrow e^+ e^-)}
\nn \\
&=& 0.13 - 0.18\,,
\end{eqnarray}
where $\mu\sim 1$ GeV. The range of values has not changed much from Eq.~(\ref{ratioasymptotic}); however, theoretical errors are reduced: corrections to Eq.~(\ref{ratioimproved}) from the $b_q^{(1)}$ and higher-order terms in Eq.~(\ref{resum}) are estimated to be roughly $40\%$, while corrections to the infinite mass limit from higher order terms are estimated to be roughly $80\%$. Corrections to Eq.~(\ref{nologs}) are hard to estimate; however, the range of values obtained give a rough upper limit on the ratio of branching ratios. In addition there are theory errors from neglecting higher-order terms in the perturbative expansion, as well as in the expansions in $v$ and $\eta$.
Our prediction can be compared to the measured value of $0.06\pm 0.03$, using the measurements $B(\Upsilon\rightarrow\gamma f_2) = (8\pm 4)\times 10^{-5}$ and $B(J/\psi\rightarrow\gamma f_2) = (1.38\pm 0.14)\times 10^{-3}$ \cite{Eidelman:2004wy}. Given the theoretical errors we can only conclude that our prediction does not disagree with data.

Our predictions for the ratios of $\Upsilon$ and $J/\psi$ branching fractions to $\gamma f_2$ are consistent with the results of Refs.~\cite{Baier:1,Baier:1985wv,Ma:2001tt} at leading order in the twist expansion used therein. In particular, we reproduce the suppression of the helicities $\abs{\lambda} = 1,2$ in the final state relative to $\lambda = 0$. In contrast with Ref.~\cite{Ma:2001tt}, we extract the NRQCD color-singlet matrix elements from the leptonic decay widths of $\Upsilon$ and $J/\psi$ instead of the decay widths to light hadrons, for which corrections from color-octet contributions must be taken into account for a reliable calculation \cite{Gremm:1997dq}. The leptonic decay width, however, receives large corrections at NNLO in perturbation theory \cite{Beneke:1997jm, Czarnecki:1997vz}. In either case, one hopes that the uncertainties are mitigated in taking the ratios of branching fractions.

We can also compare the decay rates of $J/\psi$ and $\psi'$ to $\gamma f_2$ predicted by Eq.~(\ref{2body}) at the matching scale $\mu = M$, where $a_1^q(x;M) = 0$ and $a_1^g(x;M)$ is a constant. Dependence on the integral of the wave function $\phi_g^{f_2}(x;M)$ cancels out in the ratio of branching fractions:
\begin{equation}
\frac{B(J/\psi\rightarrow\gamma f_2)}{B(\psi'\rightarrow\gamma f_2)} = \frac{B(J/\psi\rightarrow e^+ e^-)}{B(\psi'\rightarrow e^+ e^-)} = 7.85\pm 0.35\,,
\end{equation}
while the measured value is $6.57\pm 1.42$. We used $B(\psi'\rightarrow e^+ e^-) = (7.55\pm 0.31)\times 10^{-3}$ and $B(\psi'\rightarrow \gamma f_2) = (2.1\pm 0.4)\times 10^{-4}$ \cite{Eidelman:2004wy}.

\subsection{Three body decay: $\Upsilon\rightarrow\gamma\pi\pi$}

Next we consider a two pion final state in the kinematic region where the pions are collinear to each other with large energy and small total invariant mass $m_{\pi\pi} \sim \Lambda$. In this case we we have a three body final state where the two pions are collinear. It is convenient to define the variables:
\begin{equation}
\mpipi = (p_1 + p_2)^2 \,, \qquad z = \frac{\bar n\mcdot p_1}{\mups} \,.
\end{equation}
In terms of these variables the differential decay rate is
\begin{eqnarray}\label{temp}
\frac{d\Gamma}{d\mpipi\,dz} &=& \frac{1}{512\pi^3\mups^2}
\bra{\Upsilon } \psi_{{\bf p}'}^\dagger \sigma_\perp^i \chi_{-{\bf p}'}  
\chi^\dagger_{-{\bf p}} \sigma_\perp^i \psi_{\bf p} \ket{ \Upsilon} 
\nn \\ 
& &  \times  
\bigg[ \int^1_{-1} dx \big( a_1^g(x;\mu)  \phi^{\pi\pi}_g(x;\mu) + a_1^q(x;\mu)  \phi^{\pi\pi}_q(x;\mu) \big) \bigg]^2
\end{eqnarray}
to leading order in $\mpipi/\mups^2$. The properties of the meson pair light-cone wave function $\phi^{\pi\pi}$ have been investigated in Refs.~\cite{Grozin:1986at,Grozin:1983tt}, which interestingly find that in the region where $\Lambda \ll m_{\pi\pi} \ll \mups$ they are given by an integral over two single-particle wave functions. The ratio of the $\Upsilon$ and $J/\psi$ rates to $\gamma\pi\pi$ in the kinematic region of low $\mpipi$ is numerically the ratio in Eq.~(\ref{ratioimproved}) times an extra factor of $m_c^2/m_b^2 \sim 0.07 - 0.1$, that is, $0.01 - 0.02$. This suppression is due to the much larger total phase space available in $\Upsilon \to \gamma\pi\pi$ relative to that in $\Upsilon \to \gamma f_2$. No $\Upsilon\to\gamma\pi\pi$ events have yet been observed in the region $\mpipi < (1.0\text{ GeV})^2$ \cite{Anastassov:1998vs}.

\section{Conclusions}

We have systematically analyzed the exclusive radiative decays of quarkonium to energetic light mesons within the framework of soft-collinear effective theory and non-relativistic QCD to leading order in the effective theory power counting, as well as to leading order in the strong coupling. We show that color-octet contributions are suppressed by a factor of $v^2 \alpha_s(\mu_c) / \sqrt{\alpha_s(M)} \approx 0.05$ in exclusive $\Upsilon$ decays,  and can therefore be safely neglected. This is different from the situation in inclusive radiative decays in the endpoint region where octet contributions must be kept. 

We then turn to the color-singlet contribution. The tree-level matching onto this operator is carried out in Refs.~\cite{Fleming:2002rv, Fleming:2002sr}; however, the authors do not consider the complete set of operators that could appear in this decay. We use the symmetries of SCET and NRQCD, including RPI, to  show that the operator which is matched onto in Refs.~\cite{Fleming:2002rv, Fleming:2002sr} is the only operator that can appear for the decays in question. We also consider the set of possible quark  operators which can arise. Again only one of the possible quark operators can contribute to  the decays we are interested in. This operator has zero matching coefficient, but it can be generated through running.  We use the results of Ref.~\cite{Fleming:2004rk} for the renormalization group mixing of the quark and gluon operators, thus resumming large logarithms. Our results agree with the leading-twist  analyses carried out in Refs.~\cite{Baier:1,Baier:1985wv,Ma:2001tt}. In Ref.~\cite{Ma:2001tt} higher twist corrections to the color-singlet contribution were considered. These corrections would be part of higher-order SCET corrections, which we have not studied here. Such an analysis would be complicated by the subleading color-octet contribution discussed in this work. 

Finally we study the phenomenology of quarkonium radiative decay to the $f_2$, as well as to $\pi\pi$ where the pions are collinear. We make predictions for the ratios of branching fractions $B(\Upsilon\to\gamma f_2)/B(J/\psi\to\gamma f_2)$ and $B(J/\psi\to\gamma f_2)/B(\psi'\to\gamma f_2)$, as well as for the differential decay rates of $\Upsilon$ and $J/\psi$ to $\gamma\pi\pi$ in the kinematic region of two collinear pions. Our predictions for the decays to $\gamma f_2$ are consistent with experimental data, but with large theoretical uncertainties, while there is insufficient data for $\gamma\pi\pi$ with which to compare. Further theoretical work and more experimental data, especially for the light-cone wave functions of $f_2$, will improve the precision of these predictions greatly.

\acknowledgments
We thank Christian Bauer, Dan Pirjol, and Mark Wise for helpful discussions. S.F.~would like to thank the  Caltech theory group for their hospitality while part of this work was being completed. This work was supported in part by the Department of Energy under Grants No.~DE-FG03-97ER40546 (S.F.) and No.~DE-FG03-92ER40701 (C.L.), and in part by the National Science 
Foundation under Grant No.~PHY-0244599 (A.K.L.). C.L.~was supported in part by a National Defense Science and Engineering Graduate Fellowship.   A.K.L.~was supported in part by the Ralph E.~Powe Junior Faculty Enhancement Award.

\appendix

\section{Nonperturbative Matrix Elements and Light-Cone Wave Functions \label{app:lc}}

The matrix elements in Eq.~(\ref{wfdefs}) defining the SCET wave functions $\phi_{g,q}^{F_n}$ can be related to conventional QCD wave functions for flavor-singlet mesons. The two-gluon wave functions for a meson with momentum $q$ and net helicity $\lambda=0,\pm 2$ are defined as~\cite{Chernyak:1983ej}:
\begin{equation}
\begin{split}
\bra{0}\Tr G_{\mu\nu}(z)Y(z,-z)&G_{\nu\lambda}(-z)\ket{q,\lambda=0}_{\mu_0} = f_S^L q_\mu q_\lambda\int_{-1}^1 d\zeta\,e^{iz\cdot q \zeta}\phi_S^L(\zeta,\mu_0) \,, \\
\bra{0}\Tr G_{\mu\nu}(z)Y(z,-z)&G_{\nu\lambda}(-z)\ket{q,\lambda=\pm 2}_{\mu_0} \\
&= f_S^\perp[(q_\mu e^\perp_{\nu\beta} - q_\nu e^\perp_{\mu\beta})q_\alpha - (q_\mu e^\perp_{\nu\alpha} - q_\nu e^\perp_{\mu\alpha})q_\beta] 
 \int_{-1}^1 d\zeta\,e^{iz\cdot q \zeta}\phi_S^\perp(\zeta,\mu) \,,
\end{split}
\end{equation}
where $Y(z,-z)$ is the path-ordered exponential of gluon fields:
\begin{equation}
Y(z,-z) = P\exp\biggl[ig\int_{-z}^z d\sigma\cdot A(\sigma)\biggr].
\end{equation}
Going to the light-cone frame where $q_\mu = \frac{\bar n\cdot q}{2}n_\mu$ and $z^\mu = \frac{n\cdot z}{2}\bar n^\mu$, we invert these formulas to find 
\begin{equation}
\label{lightconewfs}
\begin{split}
\phi_S^L(\zeta;\mu_0) &= \frac{\bar n^\mu \bar n^\lambda}{4\pi f_S^L q^-}\int_{-\infty}^\infty dz^+\,e^{-i\zeta q^- z^+/2}\bra{0}\Tr G_{\mu\nu}(z^+)Y(z^+,-z^+){G^\nu}_\lambda(-z^+)\ket{q,\lambda=0}  \,,\\
\phi_S^\perp(\zeta;\mu_0) &= \frac{\bar n^\mu \bar n^\alpha e_\perp^{*\nu\beta}}{4\pi f_S^\perp q^-}\int_{-\infty}^\infty dz^+\,e^{-i\zeta q^- z^+/2}\bra{0}\Tr G_{\mu\nu}(z^+)Y(z^+,-z^+)G_{\alpha\beta}(-z^+)\ket{q,\lambda=\pm 2},
\end{split}
\end{equation}
where $z^+ = n\mcdot z$, and $q^-=\bar n\mcdot q$. 
Now we match the QCD fields on the right-hand side to fields in SCET:
\begin{equation}
\begin{split}
\bar n^\mu G_{\mu\nu}(z^+) &\rightarrow \left[e^{-i\bnP z^+/2}\bar n_\mu G_n^{\mu\nu}\right ]  \,,\\
Y(z^+,-z^+) &\rightarrow \left[W_n e^{i(\bnP^\dag + \bnP)z^+/2} W_n^\dag\right],
\end{split}
\end{equation}
where
\begin{equation}
G_n^{\mu\nu} = \frac{i}{g}[\mathcal{D}^\mu - ig A_{n,q}^\mu,\mathcal{D}^\nu - ig A_{n,q'}^\nu],
\end{equation}
with
\begin{equation}
i\mathcal{D}^\mu = \frac{n^\mu}{2}\bnP + \ppP^\mu + \frac{\bar n^\mu}{2}in\mcdot D_{us}.
\end{equation}
Therefore, for example, the matching between the QCD light-cone wave-function $\phi_S^L$ and the SCET operator is
\begin{eqnarray}
\phi_S^L(\zeta ; \mu_0 ) & \to & 
 \frac{\bar n^\mu \bar n^\lambda}{4\pi f_S^L q^-}\int_{-\infty}^\infty dz^+\,e^{-i\zeta q^- z^+/2}
 \bra{0} \Tr G^n_{\mu\nu}(0) W_n e^{i \bnP_+ z^+ /2} 
 W^\dagger_n {G^{n \nu}}_\lambda(0)\ket{q,\lambda=0} 
\nn \\
&  & \hspace{-15ex}
 = 
\frac{-1}{16 \pi f^L_s q^- } \sum_\omega \int_{-\infty}^\infty dz^+\, e^{-i(\zeta q^-  - \omega)z^+/2}
(q^{- } - \omega)(q^{- } + \omega)
\bra{0} \Tr B_\perp^\nu \delta_{\bnP_+ , \omega} B^\perp_\nu \ket{q,\lambda=0} \,.
\end{eqnarray}
Integrating over $z^+$, and converting from the discrete index $\omega$ to a continuous $\omega_c$ where $\zeta \equiv \omega_c / q^-$ we obtain the matching relation between the QCD and SCET light-cone wave functions
\begin{equation}
\label{wfdictionary}
\begin{split}
\phi_S^{L,\perp}(\zeta ;\mu) &\to -\frac{ q^-}{4f_S^L}(1-\zeta^2)\phi_g^{M(L,\perp)}(\zeta;\mu) \,.
\end{split}
\end{equation}
The SCET wave functions on the right-hand side are given by [cf. Eq.~(\ref{wfdefs})]:
\begin{equation}
\begin{split}
\bra{0}\bnP\Tr[B_\perp^\alpha\delta_{\omega,\bnP_+}B^\perp_\alpha]\ket{M(q)}&= (q^-)^2\phi_g^{M(L)} \,, \\
e_{\perp\alpha\beta}^*\bra{0}\bnP\Tr[B_\perp^\alpha\delta_{\omega,\bnP_+}B_\perp^\beta]\ket{M(q)}&= ( q^-)^2\phi_g^{M(\perp)} \,.
\end{split}
\end{equation}
Relations between the wave functions for the quark operator in QCD and SCET can be derived as in Ref.~\cite{Bauer:2002nz}.

\section{Spin-2 Polarization Tensors \label{sec:pol}}

The spin-2 polarization tensor for a particle of mass $m$ and three-momentum $\vect{k}$ can be built from spin-1 polarization vectors using Clebsch-Gordan coefficients to arrive at
\begin{equation}
\label{spin2pol}
\begin{split}
e^{\mu\nu}(\lambda=\pm 2) &= \frac{1}{2}
\begin{pmatrix}
0 & 0 & 0 & 0 \\
0 & 1 & \pm i & 0 \\
0 & \pm i & -1 & 0 \\
0 & 0 & 0 & 0
\end{pmatrix}  \,,\\
e^{\mu\nu}(\lambda = \pm 1) &= \mp\frac{1}{2m}
\begin{pmatrix}
0 & \abs{\vect{k}} & \pm i\abs{\vect{k}} & 0 \\
\abs{\vect{k}} & 0 & 0 & E_{\vect{k}} \\
\pm i\abs{\vect{k}} & 0 & 0 & \pm iE_{\vect{k}} \\
0 & E_{\vect{k}} & \pm iE_{\vect{k}} & 0
\end{pmatrix}  \,,\\
e^{\mu\nu}(\lambda = 0) &= \frac{1}{m^2}\sqrt{\frac{2}{3}}
\begin{pmatrix}
\vect{k}^2 & 0 & 0 & \abs{\vect{k}}E_{\vect{k}} \\
0 & -\frac{m^2}{2} & 0 & 0 \\
0 & 0 & -\frac{m^2}{2} & 0 \\
\abs{\vect{k}}E_{\vect{k}} & 0 & 0 & E_{\vect{k}}^2
\end{pmatrix}.
\end{split}
\end{equation}
%


\end{document}